\documentclass[iop,apj,twocolappendix,numberedappendix]{emulateapj}

\usepackage{natbib}
\usepackage{amsmath,amssymb}
\usepackage{apjfonts}
\usepackage{xspace}
\usepackage{rotating}
\usepackage{multirow}
\usepackage{dcolumn}

\usepackage[usenames]{xcolor}
\definecolor{mpl_blue}{HTML}{1F77B4}
\definecolor{mpl_orange}{HTML}{FF7F0E}
\definecolor{mpl_green}{HTML}{2CA02C}
\definecolor{mpl_red}{HTML}{D62728}

\renewcommand{\check}{\textcolor{orange}}

\usepackage[backref, breaklinks, plainpages=false, colorlinks=true, anchorcolor=cyan, linkcolor=mpl_red, citecolor=cyan, urlcolor=magenta, bookmarks=false]{hyperref}
\usepackage[all]{hypcap}
\usepackage[caption=false]{subfig}

\shorttitle{NANOGrav $11$-year Continuous Wave Constraints}
\shortauthors{The NANOGrav Collaboration}

\newcommand{\Msun}{M_\odot}
\newcommand{\mc}{\mathcal{M}}
\newcommand{\psrA}{PSR~J0613$-$0200}

\newcommand{\bayesephem}{\textsc{BayesEphem}}

\newcommand{\fgw}{f_\mathrm{gw}}
\newcommand{\fyr}{f_\mathrm{yr}}
\newcommand{\Ared}{\ensuremath{A_\mathrm{red}}}
\newcommand{\Tobs}{\ensuremath{T_\mathrm{obs}}}

\defcitealias{dfg+13}{NG5a}
\defcitealias{abb+14}{NG5b}
\defcitealias{abb+15}{NG9a}
\defcitealias{abb+17}{NG11a}
\defcitealias{abb+17b}{NG11b}

\begin{document}

\title{The NANOGrav 11-Year Data Set: Limits on Gravitational Waves from Individual Supermassive Black Hole Binaries}
\author{
K.~Aggarwal\altaffilmark{1,2},
Z.~Arzoumanian\altaffilmark{3},
P.~T.~Baker\altaffilmark{1,2},
A.~Brazier\altaffilmark{4,5},
M.~R.~Brinson\altaffilmark{6},
P.~R.~Brook\altaffilmark{1,2},
S.~Burke-Spolaor\altaffilmark{1,2},
S.~Chatterjee\altaffilmark{4},
J.~M.~Cordes\altaffilmark{4},
N.~J.~Cornish\altaffilmark{7},
F.~Crawford\altaffilmark{8},
K.~Crowter\altaffilmark{9},
H.~T.~Cromartie\altaffilmark{10},
M.~DeCesar\altaffilmark{$\dagger$11},
P.~B.~Demorest\altaffilmark{12},
T.~Dolch\altaffilmark{13},
J.~A.~Ellis\altaffilmark{$\dagger$1,2,14},
R.~D.~Ferdman\altaffilmark{15},
E.~Ferrara\altaffilmark{16},
E.~Fonseca\altaffilmark{17},
N.~Garver-Daniels\altaffilmark{1,2},
P.~Gentile\altaffilmark{1,2}, 
J.~S.~Hazboun\altaffilmark{$\dagger$18},
A.~M.~Holgado\altaffilmark{19},
E.~A.~Huerta\altaffilmark{19},
K.~Islo\altaffilmark{6},
R.~Jennings\altaffilmark{4},
G.~Jones\altaffilmark{20},
M.~L.~Jones\altaffilmark{1,2},
A.~R.~Kaiser\altaffilmark{1,2},
D.~L.~Kaplan\altaffilmark{6},
L.~Z.~Kelley\altaffilmark{21},
J.~S.~Key\altaffilmark{18},
M.~T.~Lam\altaffilmark{$\dagger$1,2},
T.~J.~W.~Lazio\altaffilmark{22,23},
L.~Levin\altaffilmark{24},
D.~R.~Lorimer\altaffilmark{1,2},
J.~Luo\altaffilmark{17},
R.~S.~Lynch\altaffilmark{25},
D.~R.~Madison\altaffilmark{1,2},
M.~A.~McLaughlin\altaffilmark{1,2},
S.~T.~McWilliams\altaffilmark{1,2},
C.~M.~F.~Mingarelli\altaffilmark{26},
C.~Ng\altaffilmark{9},
D.~J.~Nice\altaffilmark{11},
T.~T.~Pennucci\altaffilmark{$\dagger$1,2,27,28},
N.~S.~Pol\altaffilmark{1,2},
S.~M.~Ransom\altaffilmark{10,29},
P.~S.~Ray\altaffilmark{30},
X.~Siemens\altaffilmark{6},
J.~Simon\altaffilmark{22,23},
R.~Spiewak\altaffilmark{6,31},
I.~H.~Stairs\altaffilmark{9},
D.~R.~Stinebring\altaffilmark{32},
K.~Stovall\altaffilmark{$\dagger$12},
J.~Swiggum\altaffilmark{$\dagger$6},
S.~R.~Taylor\altaffilmark{$\dagger$22,23},
J.~E.~Turner\altaffilmark{1,2,6},
M.~Vallisneri\altaffilmark{22,23},
R.~van~Haasteren\altaffilmark{22}, 
S.~J.~Vigeland\altaffilmark{${\color{magenta}\S}\dagger$6},
C.~A.~Witt\altaffilmark{1,2},
W.~W.~Zhu\altaffilmark{33}
(The NANOGrav Collaboration)\altaffilmark{$\star$}}

\affil{$\star$Author order alphabetical by surname}
\affil{$^{1}$Department of Physics and Astronomy, West Virginia University, P.O.~Box 6315, Morgantown, WV 26506, USA}
\affil{$^{2}$Center for Gravitational Waves and Cosmology, West Virginia University, Chestnut Ridge Research Building, Morgantown, WV 26505, USA}
\affil{$^{3}$X-Ray Astrophysics Laboratory, NASA Goddard Space Flight Center, Code 662, Greenbelt, MD 20771, USA}
\affil{$^{4}$Department of Astronomy, Cornell University, Ithaca, NY 14853, USA}
\affil{$^{5}$Cornell Center for Advanced Computing, Ithaca, NY 14853, USA}
\affil{$^{6}$Center for Gravitation, Cosmology and Astrophysics, Department of Physics, University of Wisconsin-Milwaukee,\\ P.O.~Box 413, Milwaukee, WI 53201, USA}
\affil{$^{7}$Department of Physics, Montana State University, Bozeman, MT 59717, USA}
\affil{$^{8}$Department of Physics and Astronomy, Franklin \& Marshall College, P.O.~Box 3003, Lancaster, PA 17604, USA}
\affil{$^{9}$Department of Physics and Astronomy, University of British Columbia, 6224 Agricultural Road, Vancouver, BC V6T 1Z1, Canada}
\affil{$^{10}$University of Virginia, Department of Astronomy, P.O.~Box 400325, Charlottesville, VA 22904, USA}
\affil{$^{11}$Department of Physics, Lafayette College, Easton, PA 18042, USA}
\affil{$^{12}$National Radio Astronomy Observatory, 1003 Lopezville Rd., Socorro, NM 87801, USA}
\affil{$^{13}$Department of Physics, Hillsdale College, 33 E.~College Street, Hillsdale, Michigan 49242, USA}
\affil{$^{14}$Infinia ML, 202 Rigsbee Avenue, Durham, NC 27701, USA}
\affil{$^{15}$Department of Physics, University of East Anglia, Norwich, UK}
\affil{$^{16}$NASA Goddard Space Flight Center, Greenbelt, MD 20771, USA}
\affil{$^{17}$Department of Physics, McGill University, 3600  University St., Montreal, QC H3A 2T8, Canada}
\affil{$^{18}$University of Washington Bothell, 18115 Campus Way NE, Bothell, WA 98011, USA}
\affil{$^{19}$NCSA and Department of Astronomy, University of Illinois at Urbana-Champaign, Urbana, Illinois 61801, USA}
\affil{$^{20}$Department of Physics, Columbia University, New York, NY 10027, USA}
\affil{$^{21}$Center for Interdisciplinary Exploration and Research in Astrophysics (CIERA), Northwestern University, Evanston, IL 60208}
\affil{$^{22}$Jet Propulsion Laboratory, California Institute of Technology, 4800 Oak Grove Drive, Pasadena, CA 91109, USA}
\affil{$^{23}$Theoretical AstroPhysics Including Relativity (TAPIR), MC 350-17, California Institute of Technology, Pasadena, California 91125, USA}
\affil{$^{24}$Jodrell Bank Centre for Astrophysics, University of Manchester, Manchester, M13 9PL, United Kingdom}
\affil{$^{25}$Green Bank Observatory, P.O.~Box 2, Green Bank, WV 24944, USA}
\affil{$^{26}$Center for Computational Astrophysics, Flatiron Institute, 162 Fifth Avenue, New York, NY 10010, USA}
\affil{$^{27}$Institute of Physics, E\"{o}tv\"{o}s Lor\'{a}nd University, P\'{a}zm\'{a}ny P. s. 1/A, 1117 Budapest, Hungary}
\affil{$^{28}$Hungarian Academy of Sciences MTA-ELTE Extragalactic Astrophysics Research Group, 1117 Budapest, Hungary}
\affil{$^{29}$National Radio Astronomy Observatory, 520 Edgemont Road, Charlottesville, VA 22903, USA}
\affil{$^{30}$Naval Research Laboratory, Washington DC 20375, USA}
\affil{$^{31}$Centre for Astrophysics and Supercomputing, Swinburne University of Technology, PO Box 218, Hawthorn, VIC 3122, Australia}
\affil{$^{32}$Department of Physics and Astronomy, Oberlin College, Oberlin, OH 44074, USA}
\affil{$^{33}$CAS Key Laboratory of FAST, Chinese Academy of Science, Beijing 100101, China}

\affil{$^{\dagger}$ NANOGrav Physics Frontiers Center Postdoctoral Fellow}
\email[${\color{magenta}\S}$ Corresponding author email: ]{vigeland@uwm.edu}

\begin{abstract}
Observations indicate that nearly all galaxies contain supermassive black holes (SMBHs) at their centers. When galaxies merge, their component black holes form SMBH binaries (SMBHBs), which emit low-frequency gravitational waves (GWs) that can be detected by pulsar timing arrays (PTAs). We have searched the North American Nanohertz Observatory for Gravitational Waves (NANOGrav) 11-year data set for GWs from individual SMBHBs in circular orbits. As we did not find strong evidence for GWs in our data, we placed 95\% upper limits on the strength of GWs from such sources. At $\fgw = 8 \, \mathrm{nHz}$, we placed a sky-averaged upper limit of $h_0 < 7.3(3) \times 10^{-15}$. We also developed a technique to determine the significance of a particular signal in each pulsar using ``dropout'' parameters as a way of identifying spurious signals. From these upper limits, we ruled out SMBHBs emitting GWs with $\fgw = 8 \, \mathrm{nHz}$  within 120 Mpc for $\mathcal{M} = 10^9 \, M_\odot$, and within 5.5 Gpc for $\mathcal{M} = 10^{10} \, M_\odot$ at our most-sensitive sky location. We also determined that there are no SMBHBs with $\mathcal{M} > 1.6 \times 10^9 \, M_\odot$ emitting GWs with $\fgw = 2.8 - 317.8 \, \mathrm{nHz}$ in the Virgo Cluster. Finally, we compared our strain upper limits to simulated populations of SMBHBs, based on galaxies in Two Micron All-Sky Survey (2MASS) and merger rates from the Illustris cosmological simulation project, and found that only 34 out of 75,000 realizations of the local Universe contained a detectable source.

\end{abstract}
\keywords{
Gravitational waves --
Methods:~data analysis --
Pulsars:~general
}

\section{Introduction}
\label{sec:intro}

Pulsar timing arrays (PTAs) seek to detect gravitational waves (GWs) 
by searching for correlations in the timing observations 
of a collection of millisecond pulsars (MSPs). 
The stability of MSPs over long timescales ($\sim$ decades) makes PTAs ideal detectors for long-wavelength GWs 
(see \citealt{0264-9381-30-22-224002}).
Currently, there are three PTA experiments in operation: 
the North American Observatory for Gravitational Waves (NANOGrav; \citealt{ml13}), 
the European Pulsar Timing Array (EPTA; \citealt{dcl+16}), 
and the Parkes Pulsar Timing Array (PPTA; \citealt{h13}). 
Together these groups form the International Pulsar Timing Array (IPTA; \citealt{v+16}). 
The NANOGrav collaboration has released three data sets 
based on five years of observations (\citealt{dfg+13}; hereafter \citetalias{dfg+13}), 
nine years of observations (\citealt{abb+15}; hereafter \citetalias{abb+15}), 
and 11 years of observations (\citealt{abb+17}; hereafter \citetalias{abb+17}).

Potential GW sources in the PTA band include 
supermassive black hole binaries (SMBHBs; see \citealt{shm+04,s13c, 2018arXiv181108826B}), 
primordial GWs \citep{g05, lms+16}, 
cosmic strings and superstrings \citep{dv01,oms10,bos2018}, 
and bubble collisions during cosmological phase transitions \citep{cds2010}.
Historically, analyses have focused on the stochastic gravitational wave background (GWB) 
formed by the ensemble of a cosmic population of SMBHBs, 
as models predict that this signal is expected to be detected first \citep{rsg2015}. 
In the absence of a detection, constraints have been placed on the GWB, 
most recently with the NANOGrav 11-year data set (\citealt{abb+17b}, hereafter \citetalias{abb+17b}). 
These limits have been used 
to narrow the viable parameter space for binary evolution in dynamic galactic environments 
(e.g.~\citealt{tss2017, cms+2017, Mid+17}) 
and make statements about SMBHB population statistics (e.g.~\citealt{shkk2018}, \citealt{hss+2018}).

PTAs are also sensitive to GWs emitted from nearby individual SMBHBs 
with periods on the order of months to years, total masses of $\sim 10^{8} - 10^{10} \Msun$,
and orbital separations of $\sim 10^{-3} - 10^{-1} \, \mathrm{pc}$, depending on the total mass of the binary. 
SMBHBs that are emitting in the PTA band have nearly-constant orbital frequencies, 
and hence the GWs from these sources are referred to as ``continuous waves'' (CWs). 
However, we do account for the evolution of their orbits over the span of our observations in our analyses. 
Although there has not yet been a detection of GWs from individual sources 
with PTAs, they have already been used to place limits on the masses of candidate SMBHBs 
(e.g.~\citealt{jllw2004, schma+16}). 
Simulations predict that individual sources will be observed by PTAs within the next 10 -- 20 years 
\citep{rsg2015, mls+2017, kbh+2018}. 

In this paper, we present the results of searches for GWs from 
individual circular SMBHBs performed on the NANOGrav 11-year data set. 
This is an extension of 
\citealt{abb+14} (hereafter \citetalias{abb+14}), 
which performed a similar analysis on the NANOGrav 5-year data set. 
Our approach was also inspired by searches performed by the PPTA and EPTA. 
The first all-sky search for GWs from individual SMBHBs was performed 
by the PPTA in \citet{yhj+2010}, 
and a later analysis was published in \citet{zhw+2014}.
The most recent limits on GWs from individual SMBHBs 
comes from the EPTA \citep{bps+2016}, 
which performed both Bayesian and frequentist searches for GWs, and 
placed upper limits as a function of GW frequency and sky location.

The paper is organized as follows.  
In Sec.~\ref{sec:obs}, we review the pulsar observations and data reduction techniques used in the creation of the data sets. 
In Sec.~\ref{sec:data_analysis}, we describe the GW signal model 
and noise models used in our search pipelines. 
We also describe the Bayesian and frequentist methods and software. 
In Sec.~\ref{sec:results}, we present the results of detection searches. 
As we did not find evidence for GWs in the 11-year data set, 
we placed upper limits on the GW strain 
for $\fgw = 2.8 - 317.8 \, \mathrm{nHz}$. 
We also discuss a new analysis technique for identifying spurious signals in PTA data. 
In Sec.~\ref{sec:astro} we present limits on the distances to individual SMBHBs, 
and limits on the chirp masses of potential SMBHBs in the nearby Virgo Cluster. 
We also compare our current sensitivity to simulations of SMBHB populations, 
and estimate the expected number of detectable sources. 
We conclude in Sec.~\ref{sec:conclusions}. 
Throughout this paper, we use units where $G=c=1$.

\section{The $11$-year Data Set}
\label{sec:obs}

We analyzed the NANOGrav 11-year data set, 
which was published in \citetalias{abb+17} 
and consisted of times of arrival (TOAs) for 45 pulsars with observations made between 
2004 and 2015. 
Some of these data were previously published 
as the NANOGrav 5-year data set in \citetalias{dfg+13} 
and the NANOGrav 9-year data set in \citetalias{abb+15}. 
We briefly review the observations and data reduction techniques here -- 
further details can be found in \citetalias{abb+17}.

We made observations using two radio telescopes: 
the 100-m Robert C.~Byrd Green Bank Telescope (GBT) 
of the Green Bank Observatory in Green Bank, West Virginia; 
and the 305-m William E. Gordon Telescope (Arecibo) 
of Arecibo Observatory in Arecibo, Puerto Rico. 
Since Arecibo is more sensitive than GBT, 
all pulsars that could be observed from Arecibo 
($0^\circ < \delta < 39^\circ$) were observed with it, 
while those outside of Arecibo's declination range 
were observed with GBT. 
Two pulsars were observed with both telescopes: 
PSR~J1713+0747 and PSR~B1937+21. 
We observed most pulsars once a month. 
In addition, we started a high-cadence observing campaign in 2013, 
in which we made weekly observations 
of two pulsars with GBT (PSR~J1713+0747 and PSR~J1909$-$3744) 
and five pulsars with Arecibo 
(PSR~J0030+0451, PSR~J1640+2224, PSR~J1713+0747, PSR~J2043+1711, and PSR~J2317+1439). 
This high-cadence observing campaign was specifically designed to 
increase the sensitivity of our PTA to GWs from individual sources \citep{blf2011,cal+2014}.

In most cases, we observed pulsars at every epoch 
with two receivers at different frequencies 
in order to measure the pulse dispersion due to the interstellar medium (ISM). 
At GBT, the monthly observations used the 820 MHz and 1.4 GHz receivers. 
The weekly observations used only the 1.4 GHz receiver, 
which has a wide enough bandwidth to measure the dispersion. 
At Arecibo, four receivers were used for this project (327 MHz, 430 MHz, 1.4 GHz, and 2.3 GHz); 
each pulsar was observed with two different receivers, which were chosen 
based on the spectral index and timing characteristics of that pulsar. 
Backend instrumentation was upgraded about midway through our project. 
Initially, data at Arecibo and GBT were recorded 
using the ASP and GASP systems, respectively, which had bandwidths of 64 MHz. 
Between 2010 and 2012, we transitioned to the wideband systems PUPPI and GUPPI, 
which had bandwidths up to 800 MHz. 
Instrumental offsets between the data acquisition systems at each observatory 
were measured with high precision and were removed from the data to allow for seamless data sets 
(see \citetalias{abb+15} for details).

For each pulsar, the observed TOAs were fit to a timing model 
that described the pulsar's spin period and spin period derivative, 
sky location, proper motion, and parallax. 
The timing model also included terms describing pulse dispersion 
along the line of sight. 
Additionally, for those pulsars in binaries 
the timing model also included five Keplerian parameters that described the binary orbit, 
and additional post-Keplerian parameters that described relativistic binary effects 
if they improved the timing fit. 
In the GW analyses, we used a linearized timing model centered around the 
best-fit parameter values.

\section{Data Analysis Methods}
\label{sec:data_analysis}

PTAs are sensitive to GWs through their effect on the timing residuals. 
We can write the residuals for each pulsar $\delta t$ as
\begin{equation}
	\delta t = M \epsilon + n_\mathrm{white} + n_\mathrm{red} + s \,, \label{eq:residuals}
\end{equation}
where $M$ is the design matrix, which describes the linearized timing model, 
$\epsilon$ is a vector of the timing model parameter offsets, 
$n_\mathrm{white}$ is a vector describing white noise, 
$n_\mathrm{red}$ is a vector describing red noise, 
and $s$ is a vector of the residuals induced by a GW.
In this section, we briefly discuss the signal model, likelihood, and methods 
used in our analyses. 
These are all similar to those used in 
\citetalias{abb+14} -- 
in the discussion that follows, we emphasize areas in which this analysis differs from previous ones.

\subsection{Signal and noise models}
\label{sec:signal}

Consider a GW source whose location in equatorial coordinates 
is given by declination $\delta$ and right ascension $\alpha$. 
It is convenient to write the sky position in terms of the polar angle $\theta$ 
and azimuthal angle $\phi$, 
which are related to $\delta$ and $\alpha$ by 
$\theta = \pi/2-\delta$ and $\phi = \alpha$.
The emitted GWs can be written in terms of two polarizations:
\begin{equation}
	h_{ab}(t,\hat{\Omega}) = e_{ab}^{+}(\hat{\Omega}) \; h_+(t,\hat{\Omega}) + e_{ab}^{\times}(\hat{\Omega}) \; h_\times(t,\hat{\Omega}) \,,
\end{equation}
where $\hat{\Omega}$ is a unit vector from the GW source to the Solar System barycenter (SSB), 
$h_{+,\times}$ are the polarization amplitudes, 
and $e_{ab}^{+,\times}$ are the polarization tensors. 
The polarization tensors can be written in the SSB frame as \citep{wahlquist1987}
\begin{eqnarray}
	e_{ab}^{+}(\hat{\Omega}) &=& \hat{m}_a \; \hat{m}_b - \hat{n}_a \; \hat{n}_b \,, \\
	e_{ab}^{\times}(\hat{\Omega}) &=& \hat{m}_a \; \hat{n}_b + \hat{n}_a \; \hat{m}_b \,,
\end{eqnarray}
where
\begin{eqnarray}
	\hat{\Omega} &=& -\sin\theta \cos\phi \; \hat{x} - \sin\theta \sin\phi \; \hat{y} - \cos\theta \; \hat{z} \,, \\
	\hat{m} &=& -\sin\phi \; \hat{x} + \cos\phi \; \hat{y} \,, \\
	\hat{n} &=& -\cos\theta \cos\phi \; \hat{x} - \cos\theta \sin\phi \; \hat{y} + \sin\theta \; \hat{z} \,.
\end{eqnarray}
The response of a pulsar to the source is described by the antenna pattern functions $F^+$ and $F^\times$ \citep{sv2010,esc2012,thg+2016},
\begin{eqnarray}
	F^+(\hat{\Omega}) &=& \frac{1}{2} \frac{(\hat{m} \cdot \hat{p})^2 - (\hat{n} \cdot \hat{p})^2}{1+\hat{\Omega} \cdot \hat{p}} \,, \\
	F^\times(\hat{\Omega}) &=& \frac{(\hat{m} \cdot \hat{p}) (\hat{n} \cdot \hat{p})}{1+\hat{\Omega} \cdot \hat{p}} \,,
\end{eqnarray}
where $\hat{p}$ is a unit vector pointing from the Earth to the pulsar.

The effect of a GW on a pulsar's residuals can be written as
\begin{equation}
	s(t, \hat{\Omega}) = F^+(\hat{\Omega}) \; \Delta s_+(t) + F^\times(\hat{\Omega}) \; \Delta s_\times(t) \,,
\end{equation}
where $\Delta s_{+,\times}$ is the difference between the signal induced at the Earth and at the pulsar 
(the so-called ``Earth term'' and ``pulsar term''), 
\begin{equation}
	\Delta s_{+,\times}(t) = s_{+,\times}(t_p) - s_{+,\times}(t) \,,
\end{equation}
where $t$ is the time at which the GW passes the SSB and $t_p$ is the time at which it passes the pulsar. 
From geometry, we can relate $t$ and $t_p$ by
\begin{equation}
	t_p = t - L (1 + \hat{\Omega} \cdot \hat{p}) \,,
	\label{eq:pulsar_time}
\end{equation}
where $L$ is the distance to the pulsar.

For a circular binary, at zeroth post-Newtonian (0-PN) order, $s_{+,\times}$ is given by \citep{wahlquist1987,lwk+2011,cc2010}
\begin{eqnarray}
	s_+(t) &=& \frac{\mc^{5/3}}{d_L \, \omega(t)^{1/3}} \left[ -\sin 2\Phi(t) \, \left(1+\cos^2i\right) \, \cos2\psi \right. \nonumber \\
			&& \left. - 2 \cos 2\Phi(t) \, \cos i \, \sin 2\psi \right] \,, \label{eq:signal1} \\
	s_\times(t) &=& \frac{\mc^{5/3}}{d_L \, \omega(t)^{1/3}} \left[ -\sin 2\Phi(t) \, \left(1+\cos^2i\right) \, \sin2\psi \right. \nonumber \\
			&& \left. + 2 \cos 2\Phi(t) \, \cos i \, \cos 2\psi \right] \,, \label{eq:signal2}
\end{eqnarray}
where $i$ is the inclination angle of the SMBHB, $\psi$ is the GW polarization angle, 
$d_L$ is the luminosity distance to the source, 
and $\mc \equiv (m_1 m_2)^{3/5}/(m_1+m_2)^{1/5}$ 
is a combination of the black hole masses $m_1$ and $m_2$ 
called the ``chirp mass.'' 
Note that the variables $\mc$ and $\omega$ are the observed redshifted values, 
which are related to the rest-frame values $\mc_r$ and $\omega_r$ according to
\begin{eqnarray}
	\mc_r &=& \frac{\mc}{1+z} \,, \\
	\omega_r &=& \omega (1+z) \,.
\end{eqnarray}
Currently PTAs are only sensitive to sources in the local Universe for which $(1+z) \approx 1$.

For a circular binary, 
the orbital angular frequency is related to the GW frequency by $\omega_0 = \pi f_\mathrm{gw}$, 
where $\omega_0 = \omega(t_0)$. 
For our search, we defined the reference time $t_0$ as 31 December 2015 (MJD 57387), 
which corresponded to the last day data were taken for the 11-year data set. 
The orbital phase and frequency of the SMBHB are given by \citepalias{abb+14}
\begin{eqnarray}
	\Phi(t) &=& \Phi_0 + \frac{1}{32} \mc^{-5/3} \left[ \omega_0^{-5/3} - \omega(t)^{-5/3} \right] \,, \\
	\omega(t) &=& \omega_0 \left( 1 - \frac{256}{5} \mc^{5/3} \omega_0^{8/3} t \right)^{-3/8} \,,
\end{eqnarray}
where $\Phi_0$ and $\omega_0$ are the initial orbital phase and frequency, respectively. 
Unlike in \citetalias{abb+14}, we used the full expression for $\omega(t)$ in our signal model 
rather than treating the GW frequency at the Earth as a constant, 
as high-chirp-mass binaries will evolve significantly over the timescale of our observations. 

Our noise model for individual pulsars included both white noise and red noise. 
We used the same white noise model as \citetalias{abb+14}, 
which has three parameters: EFAC, EQUAD, and ECORR. 
The EFAC parameter scales the TOA uncertainties, 
and the EQUAD parameter adds white noise in quadrature. 
The ECORR parameter describes additional white noise 
added in quadrature that is 
correlated within the same observing epoch, 
such as pulse jitter \citep{dlc+2014, lcc+2017}. 
We used the improved implementation of ECORR 
described in \citetalias{abb+17b}. 
To model the red noise, we divided the noise spectrum into 30 bins spaced linearly 
between $f = 1/\Tobs$ and $f = 30/\Tobs$, 
where \Tobs\ is the total observation time 
for a particular pulsar,\footnote{\citet{vv2014} introduced a better method for choosing the frequency basis for red noise, 
which reduces the computational cost. However, in this work chose a linear frequency basis to make 
it easier to compare these results with the results in \citetalias{abb+14}.} 
and then fit the power spectral density (PSD) to a power-law model,
\begin{equation}
	P(f) = \Ared^2 \left(\frac{f}{\fyr}\right)^{-\gamma} \,,
\end{equation}
where $\fyr \equiv 1/(1 \; \mathrm{yr})$, \Ared\ is the amplitude, and $\gamma$ is the spectral index. 
There are many possible sources of red noise in pulsar timing residuals, 
including spin noise, 
variations in pulse shape, pulsar mode changes, 
and errors in modeling pulse dispersion from the ISM 
\citep{0264-9381-30-22-224002, lcc+2017, jml+2017}.
We model time-variations in the ISM through DMX parameters, 
which measure the dispersion at almost every observing epoch 
(\citetalias{abb+15}, \citealt{lcc+2016}).

\subsection{Bayesian methods and software}

We used Bayesian inference to determine posterior distributions 
of GW parameters from our data. 
The procedure followed closely that of \citetalias{abb+14}, 
with the addition of the \bayesephem\ model for the uncertainty in the SSB 
introduced in \citetalias{abb+17b}. 
Pulsar timing uses a Solar System ephemeris (SSE) to transform 
from individual observatories' reference frames 
to an inertial reference frame centered at the SSB. 
We used DE436 \citep{de436} 
to perform this transformation, 
plus the \bayesephem\ model. 
Uncertainty in the SSE has a significant impact 
on the computation of GW upper limits from PTA data. 
The \bayesephem\ model mitigates this by 
marginalizing over perturbations in the outer planets' masses and Jupiter's orbit. 
This approach removes systematic uncertainty in the position of the SSB 
by introducing statistical uncertainty through the addition of new parameters. 
Another approach to differentiating between GW signals and uncertainty in the SSE, 
which we do not explore in this paper, 
is to exploit the fact that the two have different spatial correlations \citep{thk+2016}. 
A detailed analysis of how errors in the SSE effect PTAs can be found in \citet{cgl+2018}.

We used the same likelihood as in \citetalias{abb+14}. 
We implemented the likelihood and priors and performed the searches 
using NANOGrav's new software package 
\texttt{enterprise}.\footnote{\href{https://github.com/nanograv/enterprise}{https://github.com/nanograv/enterprise}} 
We confirmed the accuracy of this package by also performing some searches using the software package 
\texttt{PAL2},\footnote{\href{https://github.com/jellis18/PAL2}{https://github.com/jellis18/PAL2}} 
which has been used for previous NANOGrav GW searches. 
Both packages used the Markov Chain Monte Carlo (MCMC) sampler 
\texttt{PTMCMCSampler}\footnote{\href{https://github.com/jellis18/PTMCMCSampler}{https://github.com/jellis18/PTMCMCSampler}} to explore the parameter space.

For detection and upper-limit runs, we described the Earth-term contribution 
to the GW signal by eight parameters: 
\begin{equation}
	\lambda_0 = \left\{ \theta, \phi, \Phi_0, \psi, i, \mc, \fgw, h_0 \right\} \,,
\end{equation}
where the characteristic strain $h_0$ is related to $\mc$, $\fgw$, and $d_L$ according to
\begin{equation}
	h_0 = \frac{2 \mc^{5/3} (\pi \fgw)^{2/3}}{d_L} \,.
	\label{eq:h0}
\end{equation}
We used log-uniform priors on $h_0$ for detection analyses, 
and a uniform prior on $h_0$ to compute upper limits on the strain.
For both types of analyses, we searched over $\log_{10} h_0 \in [-18, -11]$.

We used isotropic priors on the sky position of the source $(\theta,\phi)$, 
source inclination angle $i$, GW polarization angle $\psi$, and GW phase $\Phi_0$. 
We searched over $\log_{10} \mc$ with a uniform prior $\log_{10}(\mc/\Msun) \in [7, 10]$. 
For high $\fgw$, we truncated the prior on $\log_{10} \mc$ to account for the fact that 
high-chirp-mass systems will have merged before emitting high-frequency GWs. 
Assuming binaries merge when the orbital frequency 
is equal to the innermost stable circular orbit (ISCO) frequency, 
$\mc$ must satisfy
\begin{equation}
	\mc \leq \frac{1}{6^{3/2} \pi \fgw} \left[ \frac{q}{(1+q)^2} \right]^{3/5} \,,
\end{equation}
where $q$ is the mass ratio. 
For our analyses, we used the chirp-mass cutoff with $q=1$. 
This change to the prior on $\mc$ only affected $\fgw \geq 191.3 \, \mathrm{nHz}$.

We performed searches at fixed values of $\fgw$. 
The minimum GW frequency was set by the total observation time, 
$\fgw = 1/(11.4 \; \mathrm{yrs}) = 2.8 \, \mathrm{nHz}$. 
The maximum GW frequency was set by the observing cadence. 
Because of the high-cadence observing campaign, 
the 11-year data set can detect GWs with frequencies up to $826.7 \, \mathrm{nHz}$; 
however, the data are not very sensitive at high frequencies. 
Also, we do not expect to find any SMBHBs with orbital periods of weeks 
because high-chirp-mass systems would have already merged before emitting at those frequencies, 
and low-chirp-mass systems would be evolving through the PTA band very quickly at that point. 
Therefore, we only searched for GWs with frequencies up to $317.8 \, \mathrm{nHz}$, 
which corresponded to the high-frequency-cutoff adopted in \citetalias{abb+14}.

The pulsar-term contributions to the GW signal used the pulsar distances 
to compute the light-travel-time between when the GW passed the pulsars 
and when it passed the SSB (see~Eq.~\eqref{eq:pulsar_time}). 
We used a Gaussian prior on the distances 
with the measured mean and uncertainty from \citet{vwc+2012}; 
for the pulsars not included in that paper, 
we used a mean of 1 kpc and error of 20\%. 
The use of approximate distances for some pulsars 
did not seem to affect our results -- 
we found that the pulsar distance posteriors were identical to the priors, 
indicating that the data are insensitive to the pulsar distances. 
Furthermore, we only used approximate pulsar distances for 
pulsars that had only been observed for a few years, 
and therefore did not contribute significantly to the GW sensitivity.
The phase at the pulsar can be written as
\begin{equation}
	\Phi(t) = \Phi_0 + \Phi_p + \frac{1}{32} \mc^{-5/3} \left[ \omega(t_{p,0})^{-5/3} - \omega(t_p)^{-5/3} \right] \,,
\end{equation}
where $\Phi_p$ is the phase difference between the Earth and the pulsar. 
The pulsar phase parameters $\Phi_p$ can be computed from the 
pulsar distances and chirp mass as
\begin{equation}
	\Phi_p = \frac{1}{32} \mc^{-5/3} \left[ \omega_0^{-5/3} - \omega(t_{p,0})^{-5/3} \right] \;;
\end{equation}
however, in most cases the pulsar distance uncertainties ($\Delta L \sim 10 - 100 \; \mathrm{pc}$) 
are significantly greater than the GW wavelengths 
($\lambda_\mathrm{gw} \sim 0.1 - 10 \; \mathrm{pc}$), 
and so the phase differences between the Earth terms and pulsar terms are effectively random.
Therefore, following the approach of \citet{cc2010}, 
we treated $\Phi_p$ as an independent parameter with a uniform prior $\Phi_p \in [0, 2\pi]$.

We fixed the white noise parameters to their best-fit values, 
as determined from noise analyses performed on individual pulsars. 
In the GW analyses, we simultaneously searched over the individual pulsars' red noise 
using a power-law model 
with uniform priors on $\log_{10}\Ared \in [-20,-11]$ and $\gamma \in [0,7]$. 
In order to burn-in the red noise and \bayesephem\ parameters efficiently, 
we introduced jump proposals that drew proposed samples 
from empirical distributions based on the posteriors from an initial Bayesian analysis 
with only the pulsars' red noise and \bayesephem\ (i.e., excluding a GW signal). 
For more details, see Appendix~\ref{sec:appendix}.

We computed Bayes factors for the presence of a GW signal 
using the Savage-Dickey formula \citep{d71},
\begin{equation}
	\mathcal{B}_{10} \equiv \frac{\text{evidence}[\mathcal{H}_1]}{\text{evidence}[\mathcal{H}_0]} = \frac{p(h_0 = 0|\mathcal{H}_1)}{p(h_0 = 0|\mathcal{D},\mathcal{H}_1)} \,,
\end{equation}
where $\mathcal{H}_1$ is the model with a GW signal plus individual pulsar red noise, 
$\mathcal{H}_0$ is the model with only individual pulsar red noise, 
$p(h_0=0 | \mathcal{H}_1)$ is the prior volume at $h_0 = 0$, 
and $p(h_0 = 0 | \mathcal{D},\mathcal{H}_1)$ is the posterior volume at $h_0 = 0$. 
We were able to use the Savage-Dickey formula 
because $\mathcal{H}_1$ and $\mathcal{H}_0$ are nested models, 
i.e., $\mathcal{H}_0$ is $\mathcal{H}_1: h_0 = 0$. 
We approximated $p(h_0 = 0|\mathcal{D},\mathcal{H}_1)$ as the 
fraction of quasi-indepdent samples in the lowest-amplitude bin of a histogram of $h_0$. 
We found the quasi-independent samples by thinning the chain 
by the autocorrelation chain length, 
which is a measure of how far apart two samples in the chain must be in order to be 
statistically independent. 
We computed the error in the Bayes factor as
\begin{equation}
	\sigma = \frac{\mathcal{B}_{10}}{\sqrt{n}} \,,
\end{equation}
where $n$ is the number of samples in the lowest-amplitude bin. 

For upper limits, following the approach of \citetalias{abb+17b}, we computed the standard error as
\begin{equation}
	\sigma = \frac{\sqrt{x (1-x)/N_s}}{p(h_0=h_0^{95\%}|\mathcal{D})} \,,
\end{equation}
where $x=0.95$ and $N_s$ is the number of effective samples in the chain.
This definition of $\sigma$ is the error in the computed 95\% upper limit 
due to using a finite number of samples. 
We estimated the number of effective samples by dividing 
the total number of samples by the autocorrelation chain length, 
which is a measure of how far apart two samples in the chain must be in order to be 
statistically independent. 

\subsection{$\mathcal{F}_p$-statistic}

As in \citetalias{abb+14}, we also performed a frequentist analysis 
with the $\mathcal{F}_p$-statistic, 
which we computed using the software package \texttt{enterprise}.
The $\mathcal{F}_p$-statistic is an incoherent detection statistic 
that is derived by maximizing the log of the likelihood ratio \citep{esc2012}. 
Essentially, it is the weighted sum of the power spectrum of the residuals, 
summed over all pulsars. This statistic assumes the SMBHB's orbital frequency 
is not evolving significantly over the timescale of our observations. 
In the absence of a signal, 
$2 \mathcal{F}_p$ follows a chi-squared distribution 
with $2N_p$ degrees of freedom, where $N_p$ is the number of pulsars. 
The corresponding false-alarm-probability (FAP) is
\begin{equation}
	P_F(\mathcal{F}_{p,0}) = \exp(-\mathcal{F}_{p,0}) \sum_{k=0}^{N_p-1} \frac{\mathcal{F}^k_{p,0}}{k!} \,.
	\label{eq:Fpstat_FAP}
\end{equation}
In performing GW searches over our entire frequency range, 
we compute the $\mathcal{F}_p$ statistic $N_f$ times, 
where $N_f$ is the number of independent frequencies, 
i.e., the number of frequencies separated by $1/T_\mathrm{obs} = 2.7 \, \mathrm{nHz}$. 
The FAP for the entire search is
\begin{equation}
	P_F^T(\mathcal{F}_{p,0}) = 1 - \left[ 1 - P_F(\mathcal{F}_{p,0}) \right]^{N_f} \,.
	\label{eq:Fpstat_FAP_total}
\end{equation}
For the analysis in this paper, $N_f = 115$.

\section{Results}
\label{sec:results}

In this section we report the results of both detection and upper limit analyses 
of the NANOGrav 11-year data set for GWs from individual circular SMBHBs. 
We used the data to place upper limits as a function of frequency and sky location, 
and to compare upper limits from the 11-year data set to those from the 5- and 9-year data sets. 
We also discuss a new Bayesian technique to determine how much each pulsar 
in a PTA contributes to a common signal in order to diagnose spurious signals. 
Following the approach of \citetalias{abb+17b}, 
our analyses of the 11-year data set only used the 34 pulsars which 
had been observed for at least three years. 
Our analyses of the 5- and 9-year data sets 
used the same subset of pulsars 
that were used in the corresponding analyses for the GWB 
(\citetalias{dfg+13}, \citealt{abb+16}), 
which included 17 and 18 pulsars, respectively.

\subsection{Detection analyses}

\begin{figure}[tb]
 \centering
 \includegraphics[width=\columnwidth]{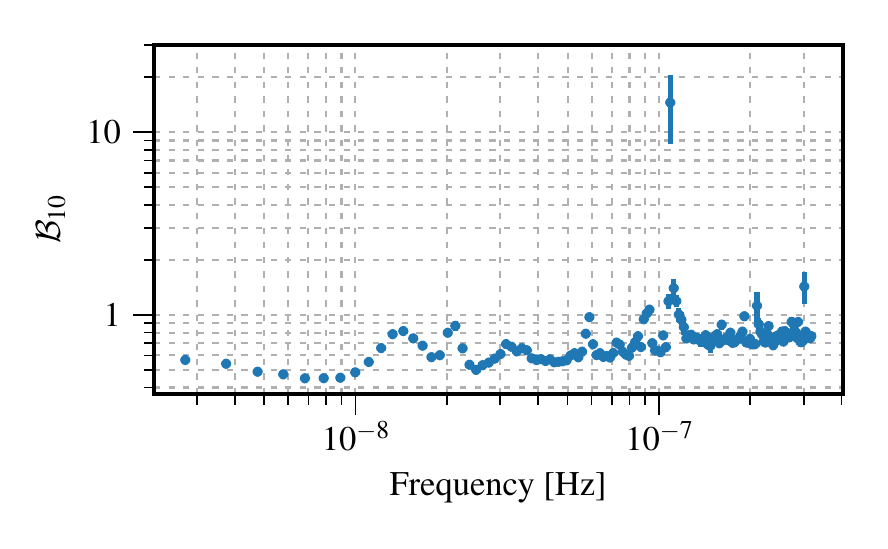}
  \caption{Bayes factors for a GW signal from an individual circular SMBHB 
		as a function of GW frequency 
  		in the NANOGrav 11-year data set. 
		We found no strong evidence for GWs in our data. 
		The highest Bayes factor was at $\fgw = 109 \, \mathrm{nHz}$, 
		for which $\mathcal{B}_{10} = 15(6)$. 
		For all other frequencies searched, the Bayes factors were close to 1.} 
  \label{fig:bayesfactors}
\end{figure}

We performed detection searches for GWs from individual circular SMBHBs 
on the 11-year data set. 
Figure~\ref{fig:bayesfactors} shows the Bayes factors for each frequency, 
marginalized over the sky location. 
We did not find strong evidence for GWs in the 11-year data set. 
The largest Bayes factor was at $\fgw = 109 \, \mathrm{nHz}$, 
for which $\mathcal{B}_{10} = 15(6)$. 
For all other frequencies, the Bayes factors were between 
$\mathcal{B}_{10} = 0.449(4)$ and $\mathcal{B}_{10} = 1.4(3)$, 
indicating no evidence of GWs in the data.

\begin{figure}[tb]
 \centering
 \includegraphics[width=\columnwidth]{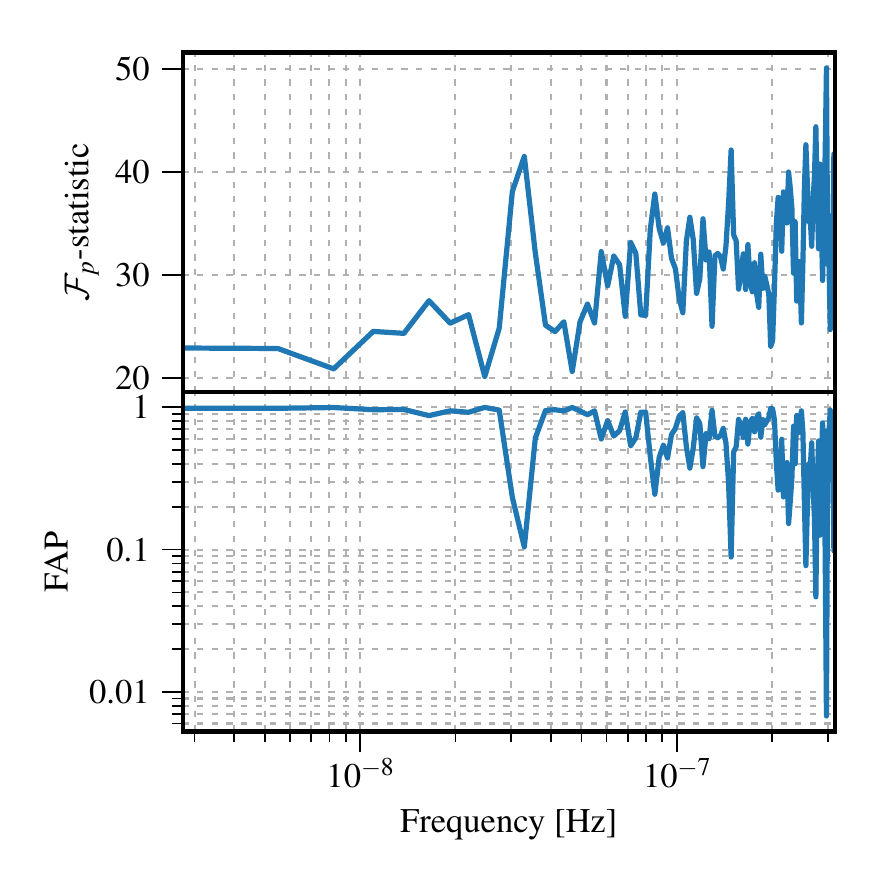}
  \caption{$\mathcal{F}_p$-statistic (top panel) and the corresponding FAP (bottom panel) for 
  		$\fgw = 2.8 - 317.8 \, \mathrm{nHz}$. 
		There were no frequencies for which the FAP was below our detection threshold of $10^{-4}$; 
		therefore, we concluded there was no evidence for GWs.}
  \label{fig:Fpstat}
\end{figure}

We also used the $\mathcal{F}_p$-statistic to determine the significance of a GW signal. 
Figure~\ref{fig:Fpstat} shows the $\mathcal{F}_p$-statistic 
as a function of $\fgw$, and the corresponding FAP as computed from Eq.~\eqref{eq:Fpstat_FAP}. 
There are no frequencies for which the FAP lies below our detection threshold of $10^{-4}$. 
At the GW frequency that maximizes $\mathcal{F}_p$, 
the total FAP for the search as computed from Eq.~\eqref{eq:Fpstat_FAP_total} is $P_F^T(\mathcal{F}_{p,0}) = 0.543$. 
Thus we concluded that the frequentist analyses also found that the 11-year data set 
does not contain significant evidence for GWs.

Although the detection search at $\fgw = 109 \, \mathrm{nHz}$ found a higher Bayes factor 
than any of the other values of $\fgw$, 
we emphasize that the Bayes factor is not high enough to claim a detection. 
A Bayes factor of 15 means 15:1 odds for the presence of a GW signal; 
similarly, at this frequency the $\mathcal{F}_p$-statistic corresponds to 
$P_F(\mathcal{F}_{p,0}) = 0.235$, or signal-to-noise ratio (S/N) of 1.2. 
Neither of these metrics support the claim that the data show evidence of GWs. 
Furthermore, as we discuss in more detail in Sec.~\ref{sec:dropout}, 
we determined that most of the evidence for this signal 
was in the residuals of a single pulsar, J1713+0747, 
whereas a true GW signal this strong should be seen in many pulsars.

\subsection{Upper limit analyses}

\begin{figure}[tb]
 \centering
 \includegraphics[width=\columnwidth]{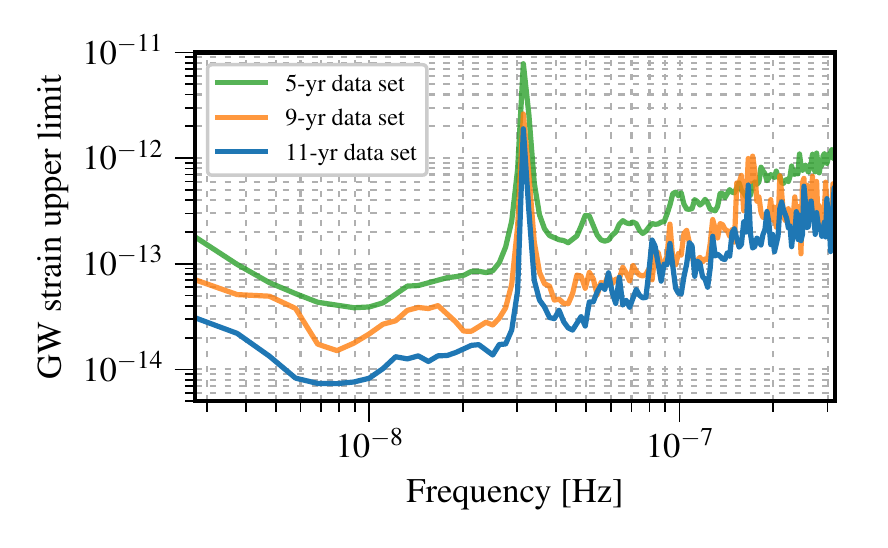}
  \caption{Sky-averaged 95\% upper limit on the GW strain amplitude 
  		as a function of GW frequency from the 
		NANOGrav 5-year data set (green), 
		9-year data set (orange), and 
		11-year data set (blue). 
		These analyses used \bayesephem\ to parametrize uncertainty in the SSB. 
		The data were most sensitive at $\fgw = 8 \, \mathrm{nHz}$, 
		with a strain upper limit of approximately 
		$h_0 < 1.51(7) \times 10^{-14}$ from the 9-year data set, 
		and $h_0 < 7.3(3) \times 10^{-15}$ from the 11-year data set.}
  \label{fig:freq_ul_9yr11yr}
\end{figure}

As we did not find strong evidence for GWs from individual circular SMBHBs in the 11-year data set, 
we placed upper limits on the GW strain. 
Figure~\ref{fig:freq_ul_9yr11yr} shows the sky-averaged 95\% upper limit on the GW strain amplitude. 
At the most sensitive frequency of $8 \, \mathrm{nHz}$, 
we placed a 95\% upper limit on the strain of approximately 
$h_0 < 7.3(3) \times 10^{-15}$. 
We also show the strain upper limits from the 5- and 9-year data sets for comparison.
There was an improvement of about a factor of two between the 5-year and 9-year data sets, 
and more than a factor of two between the 9-year and 11-year data sets. 
Our upper limit based on the 11-year data set was about 1.4 times lower than that of $h_0 < 10^{-14}$ 
set by the EPTA based on observations of 6 pulsars observed for up to 17.7 years \citep{bps+2016,dcl+16}. 
However, a direct comparison between the EPTA results and the results in this paper is complicated 
by the fact that the analysis in \citet{bps+2016} varied both the white and red noise, whereas our analysis 
varied only the red noise and fixed the white noise. 
Our upper limit is also about a factor of 2 lower than that of $h_0 < 1.7 \times 10^{-14}$ 
set by the PPTA using their Data Release 1 \citep{zhw+2014, mhb+2013}

We note that there is an increase in the strain upper limit 
from the 9-year data set at around $\fgw = 15 \, \mathrm{nHz}$; 
however, there is not a significant Bayes factor at this frequency ($\mathcal{B}_{10} = 1.4(1)$). 
Furthermore, this ``bump'' in the spectrum is not present in the 11-year data set. 
If it were caused by a GW, the significance should have increased in the 11-year data set. 
As discussed in more detail in Sec.~\ref{sec:dropout}, 
this increase in the strain upper limit is due to an unmodeled signal in a single pulsar, \psrA.

\begin{figure}[tb]
 \centering
 \includegraphics[width=\columnwidth]{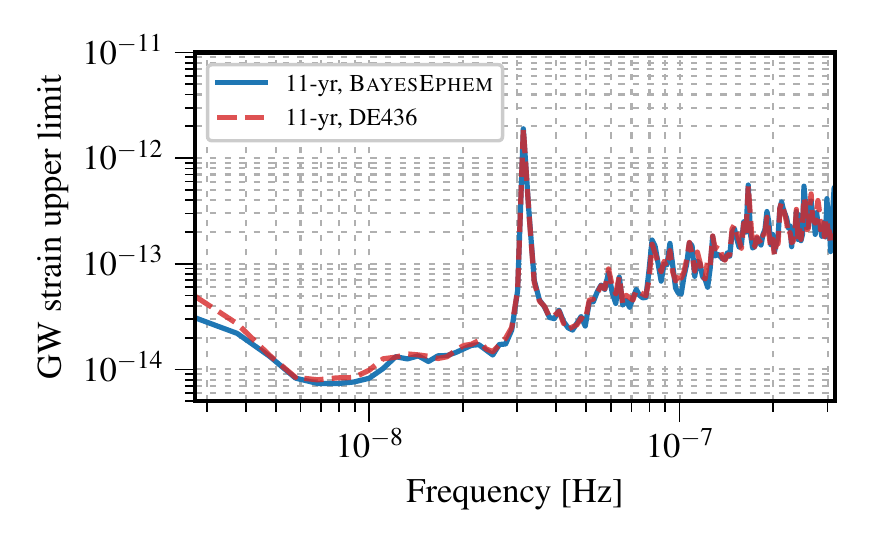}
  \caption{Sky-averaged 95\% upper limit on the 
  		GW strain amplitude from a circular SMBHB 
  		as a function of GW frequency, 
		with and without \bayesephem\ 
		(solid, blue curve and dashed, red curve, respectively). 
		At the lowest frequencies ($\fgw \lesssim 4 \; \mathrm{nHz}$), 
		the analysis with \bayesephem\ 
		was more sensitive than the analysis without, 
		but there was no difference in sensitivity at higher frequencies.}
  \label{fig:freq_ul_byephemeris}
\end{figure}

In Figure~\ref{fig:freq_ul_byephemeris}, 
we compare the sky-averaged strain upper limits computed with and without \bayesephem, 
which allows for uncertainties in the SSE. 
Including \bayesephem\ in our model resulted in a lower strain upper limit for 
$\fgw<4 \, \mathrm{nHz}$, 
but did not affect the strain upper limit at higher frequencies. 
This was expected since \bayesephem\ primarily augments the orbit of Jupiter, 
which has an orbital period of 12 years.

\begin{figure}[tb]
 \centering
 \includegraphics[width=\columnwidth]{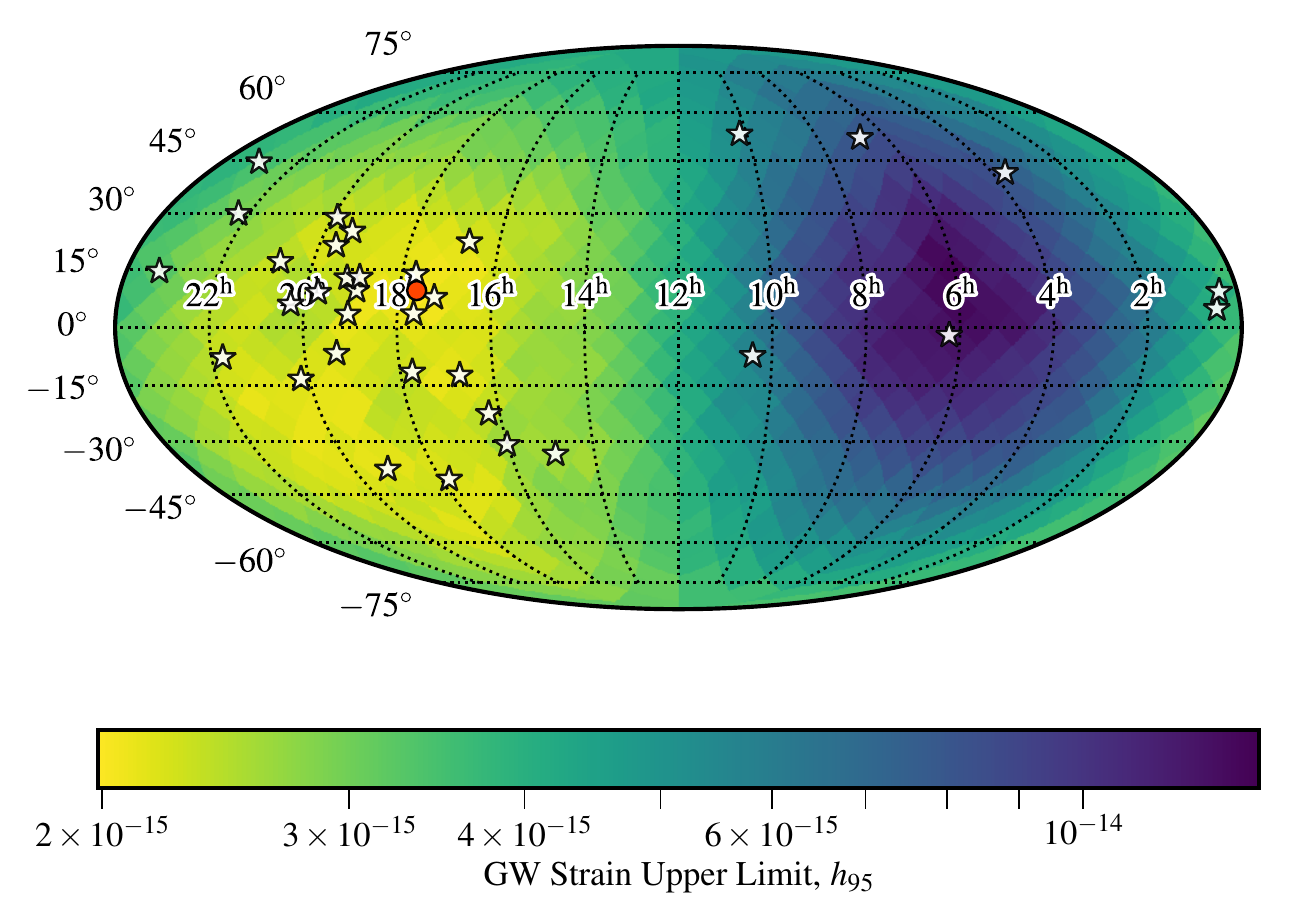}
  \caption{The 95\% upper limit on the GW strain amplitude from a circular SMBHB 
  		with $\fgw = 8 \, \mathrm{nHz}$ as a function of sky position 
  		from an analysis of the 11-year data set, plotted in equatorial coordinates using the Mollweide projection. 
		We used the DE436 ephemeris model with \bayesephem\ to model uncertainty in the SSB. 
		The positions of pulsars in our array are indicated by stars, 
		and the most sensitive sky location is indicated by a red circle. 
		The 95\% upper limit ranged from $2.0(1)\times10^{-15}$ at our most sensitive sky location 
		to $1.34(4)\times10^{-14}$ at our least sensitive sky location.}
  \label{fig:skymaps}
\end{figure}

Our sensitivity to individual sources varied significantly 
with the angular position of the source  
due to having a finite number of pulsars distributed unevenly across the sky. 
Figure~\ref{fig:skymaps} shows the 95\% upper limit on the GW strain for $\fgw = 8 \, \mathrm{nHz}$ 
as a function of sky position, plotted in equatorial coordinates. 
The upper limit varies from $h_0 < 2.0(1)\times10^{-15}$ at the most sensitive sky location 
to $h_0 < 1.34(4) \times 10^{-14}$ at the least sensitive sky location. 

\subsection{``Dropout'' analyses}
\label{sec:dropout}

Our searches of the NANOGrav 9-yr and 11-yr data sets found two low-S/N signals. 
In order to identify their sources, 
we introduced a new type of analysis that used ``dropout'' parameters 
to determine how much each individual pulsar contributed to these signals. 
In a dropout analysis, the model for a pulsar's residuals [Eq.~\eqref{eq:residuals}] 
is modified so that the GW signal could be turned on or off in each individual pulsar:
\begin{equation}
	\delta t = M \epsilon + n_\mathrm{white} + n_\mathrm{red} + \kappa \; s \,, \label{eq:dropout}
\end{equation}
where $\kappa \in \{0, 1\}$. 
The GW parameters were held fixed at the values 
that maximized the likelihood of a standard GW search, 
and dropout parameters $k_a$ were introduced into the signal model, 
which were drawn from a uniform prior between 0 and 1. 
These parameters determined whether the signal was turned on or off in a particular pulsar:
\begin{equation}
	\kappa_a = \left\{ \begin{array}{cc} 0 & k_a < k_\mathrm{threshold} \\ 
								1 & k_a \geq k_\mathrm{threshold} \end{array} \right. \,,
\end{equation}
where $k_\mathrm{threshold}$ sets the prior on whether the signal should be included 
in a pulsar. For the analyses in this paper, we used $k_\mathrm{threshold} = 1/2$, 
meaning that the prior assumed it was equally likely that the GW be turned on or off. 
At each iteration of the MCMC, the astrophysical properties of the GW were fixed, 
and the only thing that varied was which pulsars' residuals contained the GW signal. 
The posteriors of the dropout parameters indicated how much support there was for the GW 
in each pulsar.

The dropout method tests the robustness of the correlations in the signal 
by determining whether evidence for the signal comes from correlations between multiple pulsars, 
or it only originates from a single pulsar. 
It is similar to the dropout technique in neural networks, 
where units are randomly dropped during training in order to strengthen the network \citep{shk+2014}. 
This method is also similar to jackknife resampling \citep{efron1981}; 
however, in jackknifing, samples are removed in order to estimate the bias in parameter estimation, 
whereas in dropout analyses the parameter values are held fixed, and the dropout parameters 
indicate how much each pulsar is biasing the parameter estimation. 
An upcoming paper will further describe and develop this method \citep{vvt2019}

\begin{figure}[tb]
 \centering
 \includegraphics[width=\columnwidth]{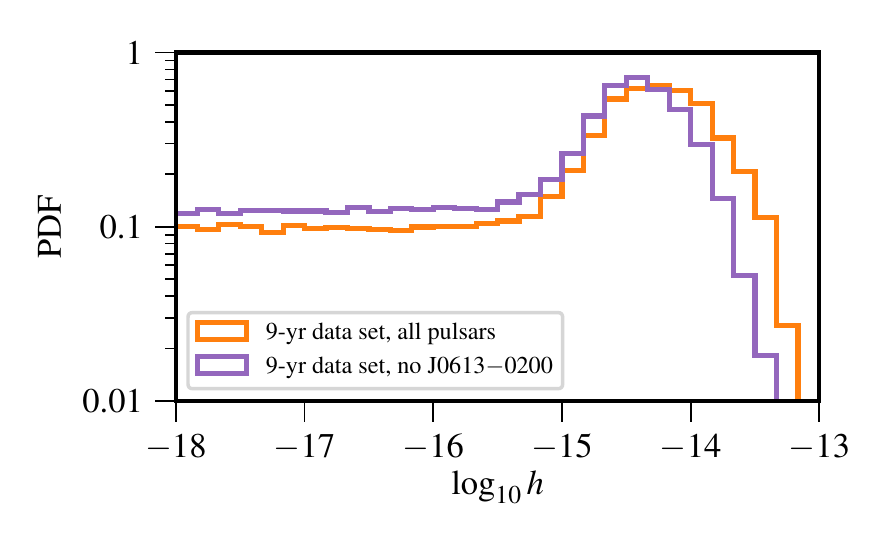}
  \caption{Comparison between a search at $\fgw = 15 \, \mathrm{nHz}$ 
  		performed on the 9-yr data set with all pulsars (orange) and excluding \psrA\ (purple). 
		There was very little difference between the Bayes factors 
		($\mathcal{B}_{10} = 1.4(1)$ with all pulsars, and $\mathcal{B}_{10} = 1.11(4)$ excluding \psrA), 
		but there was a significant difference in the 95\% strain upper limit. 
		We found an upper limit of $4.1(2) \times 10^{-14}$ with all pulsars, 
		compared with $3.2(3) \times 10^{-14}$ without \psrA.} 
  \label{fig:9yr_bump}
\end{figure}

\begin{figure}[tb]
 \centering
 \includegraphics[width=\columnwidth]{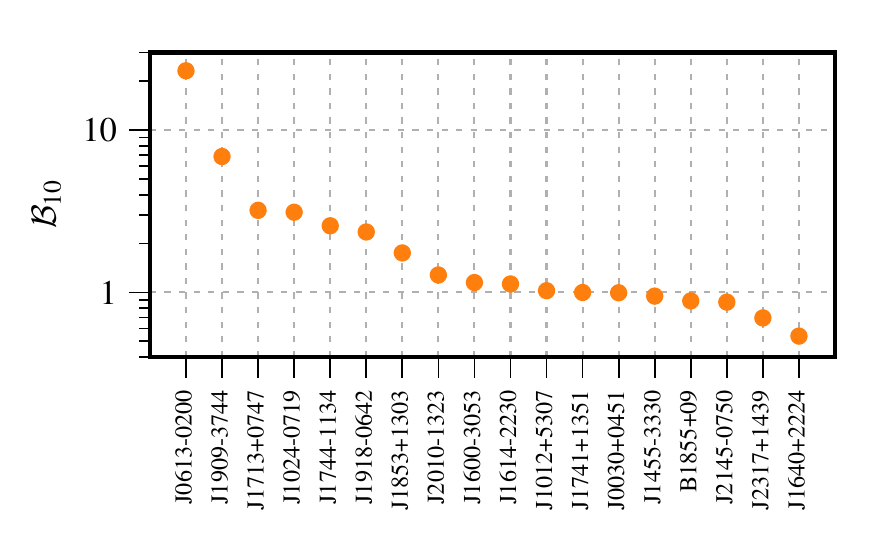}
  \caption{Bayes factors for the presence of a GW signal in each pulsar's residuals, 
  		from an analysis of the 9-yr data set with $\fgw = 15 \, \mathrm{nHz}$. 
  		The GW parameters are fixed to the maximum-likelihood values, 
		as determined from a detection analysis. 
  		\psrA\ had the largest Bayes factor for the signal, with $\mathcal{B}_{10} = 23.2(5)$, 
		indicating that \psrA\ was the primary source of this signal.}
  \label{fig:dropout}
\end{figure}

We performed two dropout analyses. 
The first was on the 9-yr data set at $\fgw = 15 \, \mathrm{nHz}$. 
The analysis of the 9-year data set found an increase 
in the 95\% strain upper limit at $\fgw = 15 \, \mathrm{nHz}$ 
compared to the upper limits at neighboring frequencies. 
Furthermore, as shown in Figure~\ref{fig:9yr_bump}, 
we found that the strain upper limit decreased significantly 
when \psrA\ was removed from the 9-year data set. 
However, there was very little difference in the Bayes factor: 
$\mathcal{B}_{10} = 1.4(1)$ with all pulsars, and $\mathcal{B}_{10} = 1.11(4)$ excluding \psrA. 
Figure~\ref{fig:dropout} shows the results of a dropout analysis. 
We fixed the GW signal parameters to the best-fit values from a detection analysis including all pulsars, 
and only allowed the dropout parameters to vary. 
We set $k_\mathrm{threshold} = 1/2$, 
so that there was an equal prior probability of the signal being included or excluded 
in the model for each pulsar's residuals. 
\psrA\ had the largest Bayes factor while all other pulsars had Bayes factors of order 1, 
from which we concluded that the increase in the strain upper limit at $\fgw = 15 \, \mathrm{nHz}$ 
was caused by an unmodeled non-GW signal in \psrA. 
We have applied advanced noise modeling techniques to this pulsar, 
using more complex models for the red noise, 
and modeling time-dependent variations in the dispersion as a Gaussian process rather than with DMX. 
These results will be discussed in an upcoming paper.

\begin{figure}[tb]
 \centering
 \includegraphics[width=\columnwidth]{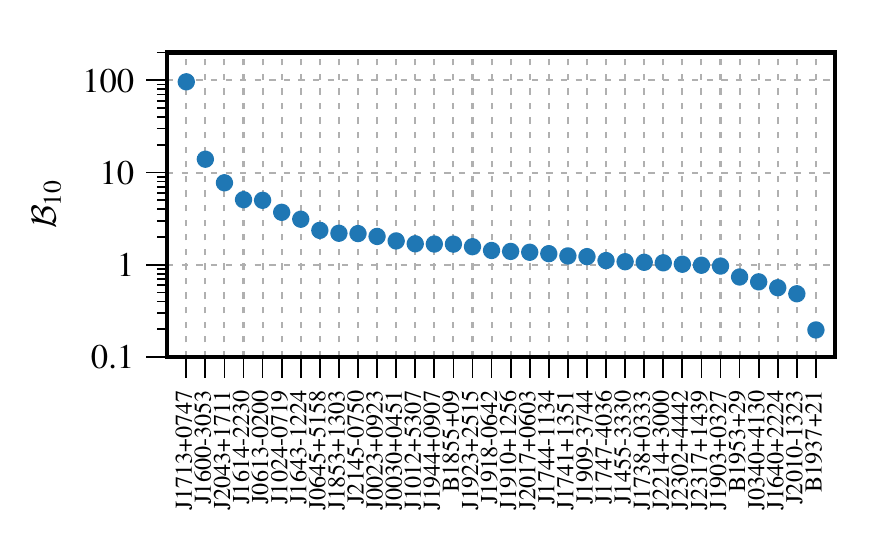}
  \caption{Bayes factors for the presence of a GW signal in each pulsar's residuals, 
  		from an analysis of the 11-yr data set with $\fgw = 109 \, \mathrm{nHz}$. 
  		The GW parameters were fixed to the maximum-likelihood values, 
		as determined from a detection analysis. 
		We concluded that this signal was primarily coming from PSR J1713+0747, 
		for which $\mathcal{B}_{10} = 96.2(1)$. 
		}
  \label{fig:dropout_gremlin}
\end{figure}

We also performed a dropout analysis on the 11-yr data set 
at $\fgw = 109 \, \mathrm{nHz}$, 
for which the detection searches had found $\mathcal{B}_{10} = 15(6)$. 
Figure~\ref{fig:dropout_gremlin} shows the Bayes factors for each pulsar's dropout parameter. 
We found that PSR J1713+0747 had the strongest Bayes factor for including a GW signal 
at this frequency, with $\mathcal{B}_{10} = 96.2(1)$, 
indicating that most of the evidence for this signal comes from the residuals of PSR J1713+0747. 
We did not perform an analysis removing PSR J1713+0747 
because it is one of the most sensitive pulsars in the NANOGrav PTA, 
and removing it always decreases our sensitivity to GWs. 
Since J1713+0747 significantly contributes to every GW analysis, 
it is unsurprising that noise in this pulsar can be confused for a GW. 
A noise analysis of J1713+0747 is underway 
using the advanced noise modeling techniques 
that were also applied to J0613$-$0200, 
and the results will be discussed in an upcoming paper. 
Future CW analyses of PTA data will be able to definitively determine 
the source of this signal with additional timing data 
and the incorporation of advanced noise modeling techniques into 
the data analysis pipeline.

\section{Limits on Astrophysical Properties of Nearby SMBHBs}
\label{sec:astro}

\begin{figure}[tb]
 \centering
 \includegraphics[width=\columnwidth]{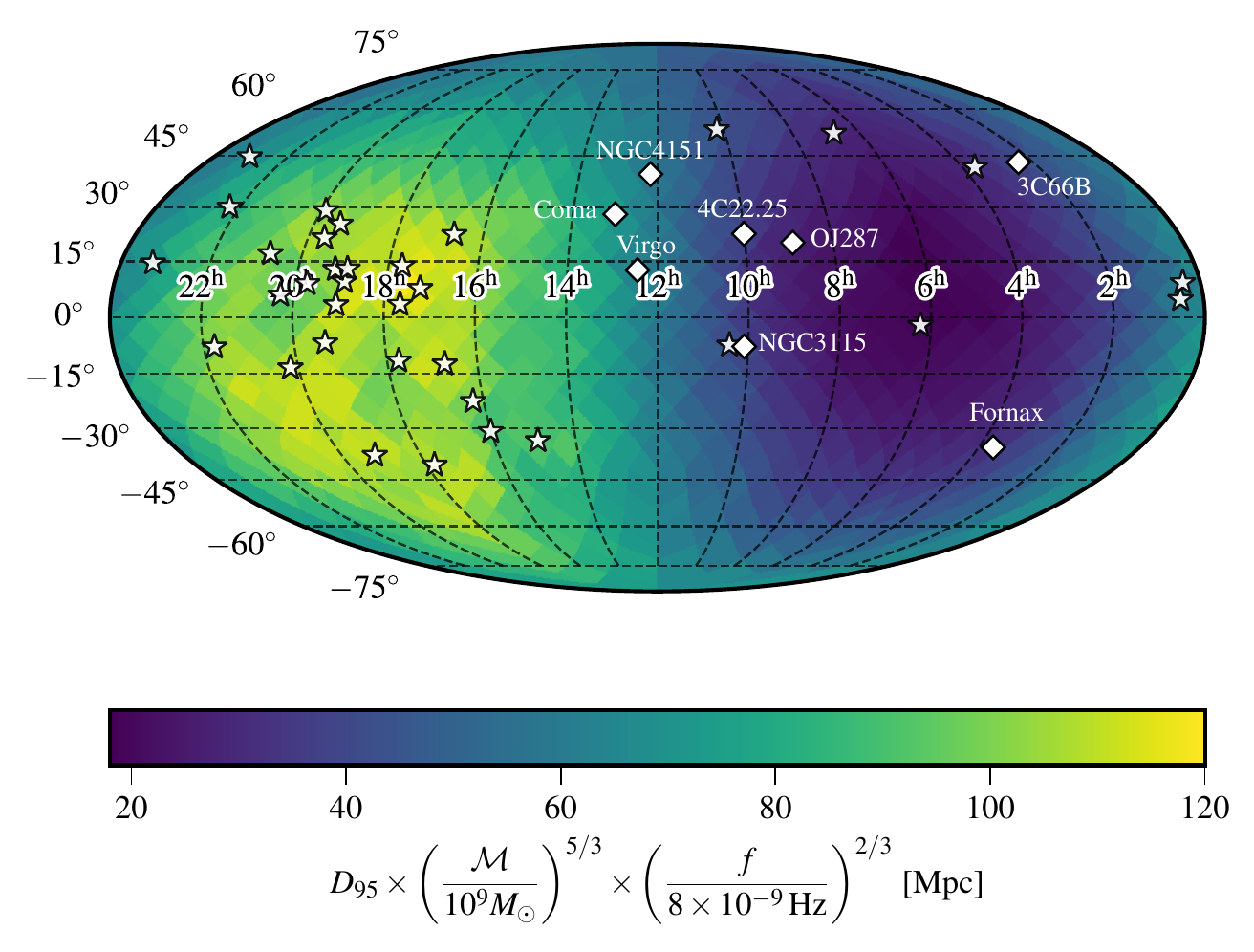}
  \caption{The 95\% lower limit on the distance to individual SMBHBs with 
  		$\mathcal{M} = 10^9 \, M_\odot$ and $\fgw = 8 \, \mathrm{nHz}$ 
		as a function of sky position based on an analysis of the 11-year data set, 
		plotted in equatorial coordinates using the Mollweide projection. 
		The stars indicate the positions of pulsars in our array, and 
		the diamonds indicate the positions of known SMBHB candidates 
		or galaxy clusters that may contain SMBHBs. 
		At our most-sensitive sky location, we place a limit of $d_L > 120 \, \mathrm{Mpc}$ 
		for SMBHBs with $\mathcal{M} = 10^9 \, M_\odot$, 
		and $d_L > 5.5 \, \mathrm{Gpc}$ for SMBHBs with $\mathcal{M} = 10^{10} \, M_\odot$.
		}
  \label{fig:luminosity_skymaps}
\end{figure}

In this section, we discuss what we can infer about the astrophysical properties of 
nearby SMBHBs from our limits on the GW strain. 
We used the 95\% upper limits on the GW strain 
to place 95\% {\it lower} limits on the distance to SMBHBs 
using Eq.~\eqref{eq:h0} for a given chirp mass. 
Figure~\ref{fig:luminosity_skymaps} shows the 
95\% lower limit on the distances to individual SMBHBs as a function of sky position, 
plotted in equatorial coordinates,  
for sources with $\mathcal{M} = 10^9 \, M_\odot$ and $\fgw = 8 \, \mathrm{nHz}$. 
The limits on the luminosity distance varied by a factor of 7 between 
the most-sensitive and least-sensitive sky locations. 
At the most-sensitive sky location, we found  
$d_L > 120 \, \mathrm{Mpc}$ for SMBHBs with $\mc = 10^{9} \, \Msun$ 
and $d_L > 5.5 \, \mathrm{Gpc}$ for SMBHBs with $\mc = 10^{10} \, \Msun$. 

\begin{figure}[tb]
 \centering
 \includegraphics[width=\columnwidth]{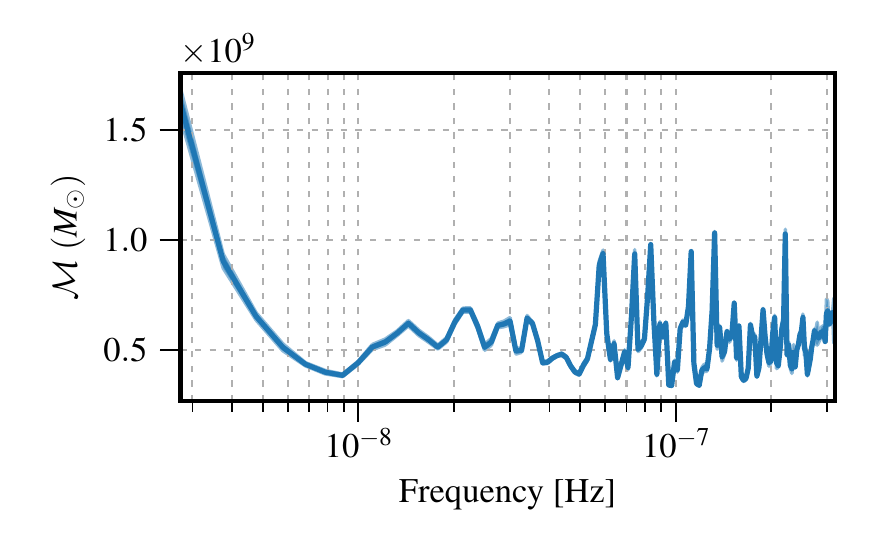}
  \caption{The 95\% upper limit on the chirp mass of any SMBHBs 
  		in the Virgo Cluster as a function of GW frequency. 
		We found that there are no SMBHBs in the Virgo Cluster with 
		$\mc > 1.6(1) \times 10^9 \, M_\odot$ emitting GWs in this frequency band. 
		At $\fgw = 9 \, \mathrm{nHz}$, we placed an upper limit of 
		$3.8(1) \times 10^8 \, M_\odot$.}
  \label{fig:Virgo_chirpmass}
\end{figure}

Figure~\ref{fig:Virgo_chirpmass} shows the limits on the chirp masses of any SMBHBs 
in the nearby Virgo Cluster, which is at a distance of $16.5 \, \mathrm{Mpc}$. 
We found that there are no SMBHBs in the Virgo Cluster with 
$\mc > 1.6(1) \times 10^9 \, M_\odot$ emitting GWs in the PTA band. 
Furthermore, there are no SMBHBs with $\mc > 3.8(1) \times 10^8 \, M_\odot$ 
emitting GWs with $\fgw = 9 \, \mathrm{nHz}$. 
These chirp-mass limits imply that none of the galaxies 
NGC 4472 (estimated black hole mass of $2.5 \times 10^9 \, M_\odot$; \citealt{rts+2013}), 
NGC 4486 (estimated black hole mass of $6.6 \times 10^9 \, M_\odot$; \citealt{gar+2011}), 
or NGC 4649 (estimated black hole mass of $4.5 \times 10^9 \, M_\odot$; \citealt{sg2010})
could contain binaries emitting GWs in this frequency range.

\begin{figure}[tb]
 \centering
 \includegraphics[width=\columnwidth]{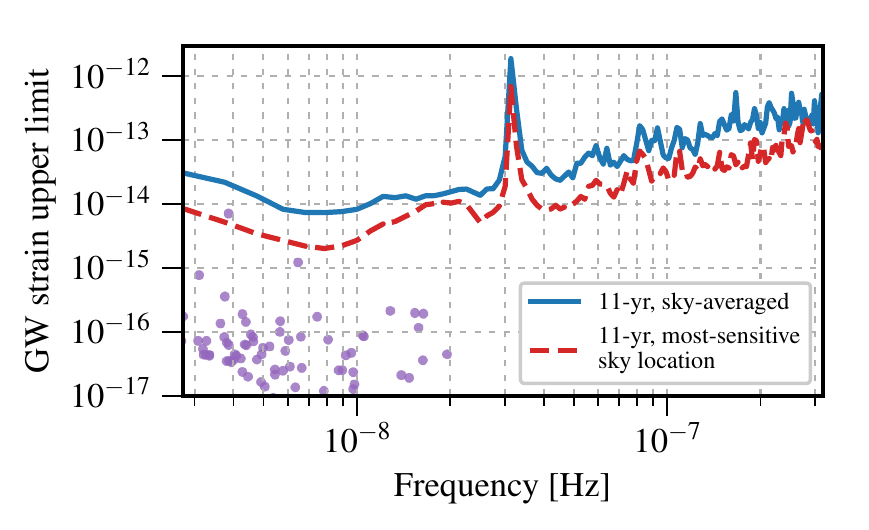}
  \caption{GW frequency and strain for the loudest GW sources for a sample realization of the local Universe, 
		plotted alongside our 95\% strain upper limit curve. 
		This simulation used simulated populations of nearby SMBHBs from \citet{mls+2017} 
		to determine the number of sources emitting GWs in the PTA band. 
		For this realization, there are 87 SMBHBs -- none of them lie above the sky-averaged strain upper limit curve, 
		and there is one source that lies above the strain upper limit curve at the most-sensitive sky location. 
		This source could be detectable depending on its sky location.}
  \label{fig:detectable_sources}
\end{figure}

In order to assess how likely we were to have detected a SMBHB given our current sensitivity, 
we compared our strain upper limit curves to simulations of nearby SMBHBs. 
A similar technique was introduced in \citet{bps+2016} 
to estimate the detection probability from the strain upper limit curve. 
We used simulated populations of SMBHBs from \citet{mls+2017}, 
which are based on galaxies in the Two Micron All-Sky Survey (2MASS; \citealt{scs+2006}) 
and merger rates from the Illustris cosmological simulation project \citep{gvs+2014,rgv+2015}. 
We estimated the number of detectable sources 
as the number lying above our sky-averaged 95\% strain upper limit curve. 
Figure~\ref{fig:detectable_sources} shows the loudest GW sources for a sample realization, 
plotted alongside our 95\% strain upper limit curve. 
We show both the sky-averaged strain upper limit curve (solid, blue line) 
and the strain upper limit curve at the most-sensitive sky location (dashed, red line). 
For this particular simulation, none of the sources were above the 
sky-averaged strain upper limit curve; 
therefore, we concluded there were no detectable sources in this particular realization. 
Out of 75,000 realizations of the local Universe, 
34 contained a source that lay above the sky-averaged strain upper limit curve 
(i.e., 0.045\% of realizations contained an observable SMBHB), 
from which we concluded that our non-detection was unsurprising given our current sensitivity. 
We point out, though, that our sensitivity varies significantly with sky location, 
and therefore some sources that are below the sky-averaged strain upper limit curve 
may be detectable depending on their sky locations. 
In our simulations, we found that a GW source lay above the strain upper limit curve 
at the most-sensitive sky location in 918 realizations (1.22\%).

\section{Summary and Conclusions}
\label{sec:conclusions}

We searched the NANOGrav 11-year data set for GWs 
from individual circular SMBHBs. 
As we found no strong evidence for GWs in our data, we placed limits on the GW strain. 
We determined that the 11-year data set was most sensitive to $\fgw = 8 \, \mathrm{nHz}$, 
for which the sky-averaged strain upper limit was $h_0 < 7.3(3) \times 10^{-15}$. 
We produced sky maps of the GW strain upper limit at $\fgw = 8 \, \mathrm{nHz}$. 
At the most sensitive sky location, we placed a strain upper limit of $h_0 < 2.0(1) \times 10^{-15}$. 
These results are the first limits on GWs from individual sources 
to be robust to uncertainties in the SSE 
due to the incorporation of \bayesephem\ in our model. 
As shown in Fig.~\ref{fig:freq_ul_byephemeris}, 
uncertainty in the SSE only affects our sensitivity to GWs for $\fgw < 4 \, \mathrm{nHz}$.

We introduced a new detection technique 
that uses ``dropout'' parameters to determine 
the significance of a common signal in each individual pulsar. 
We applied this technique to two low-S/N signals 
found in the 9-year and 11-year data sets, and 
identified the pulsars contributing the most to these signals. 
This technique is currently being used within NANOGrav 
in other GW searches, 
and a methods paper developing this technique is underway. 
Determining the physical processes causing these low-S/N signals 
is beyond the scope of this paper. 
Advanced noise analyses of all the pulsars in the NANOGrav PTA 
are underway, using more complicated models for the red noise and 
incorporating models for time-variations in the dispersion measure, 
and the methods and results will be the subject of an upcoming paper.

We used our strain upper limits to place 
lower limits on the luminosity distance to individual SMBHBs. 
At the most sensitive sky location, we placed a limit of 
$d_L > 120 \, \mathrm{Mpc}$ for $\mc = 10^{9} \, \Msun$ 
and $d_L > 5.5 \, \mathrm{Gpc}$ for $\mc = 10^{10} \, \Msun$. 
Our non-detection of GWs was not surprising given our current sensitivity limits. 
Using simulated populations of nearby SMBHBs from \citet{mls+2017}, 
we found that only 34 out of 75,000 realizations of the local Universe 
contained a SMBHB whose GW strain lay above our sky-averaged 95\% upper limit curve. 
These simulations also supported the conclusion that the two low-S/N signals 
found in the 9-year and 11-year data sets were not GW signals.

Although we have not yet made a positive detection of GWs from individual SMBHBs, 
the NANOGrav PTA is sensitive enough to place interesting limits on such sources. 
Based on our non-detection of GWs, we have determined 
that there are no SMBHBs in the Virgo Cluster with 
$\mc > 1.6(1) \times 10^{9} \, \Msun$ emitting GWs in the PTA band. 
Furthermore, our sensitivity to GWs from individual SMBHBs will continue to improve 
as we increase our observation times, add MSPs to our array, 
and develop improved pulsar noise models.

\acknowledgements

\emph{Author contributions.}

This paper is the result of the work of dozens of people over the course of more than thirteen years. 
We list specific contributions below. 
ZA, KC, PBD, MED, TD, JAE, ECF, RDF, EF, PAG, GJ, MLJ, MTL, LL, DRL, RSL, MAM, CN, DJN, TTP, SMR, PSR, RS, IHS, KS, JKS, and WZ developed the 11-year data set. 
SJV led the search and coordinated the paper writing. 
JAE, SRT, PTB, SJV, and CAW designed and implemented the Bayesian search algorithms in \texttt{enterprise}, 
and ARK, STM, SRT, and SJV implemented the $\mathcal{F}_p$-statistic in \texttt{enterprise}. 
NJC helped to uncover a bug in an early implementation of the search algorithms. 
PTB developed the jump proposals based on empirical distributions. 
SJV, KI, JET, KA, PRB, AMH, and NSP ran the searches. 
MRB, JAE, and SRT performed an initial analysis of the 9-year data set. 
CMFM, KI, TJWL, JS, and SJV did the astrophysical interpretation. 
SJV, KI, and CMFM wrote the paper and generated the plots.

\emph{Acknowledgments.}
We thank the anonymous referee for helpful comments and suggestions 
which improved this manuscript. 
The NANOGrav project receives support from National Science Foundation (NSF) 
Physics Frontier Center award number 1430284.
NANOGrav research at UBC is supported by an NSERC Discovery Grant and Discovery Accelerator Supplement and by the Canadian Institute for Advanced Research.
Portions of this research were carried out at the Jet Propulsion Laboratory, California Institute of Technology, under a contract with the National Aeronautics and Space Administration.
MV and JS acknowledge support from the JPL RTD program.
SRT was partially supported by an appointment to the NASA Postdoctoral Program at the Jet Propulsion Laboratory, administered by Oak Ridge Associated Universities through a contract with NASA. 
JAE was partially supported by NASA through Einstein Fellowship grants PF4-150120.
SBS and CAW were supported by NSF award \#1815664.
WWZ is supported by the Chinese Academy of Science Pioneer Hundred Talents Program, the Strategic Priority Research Program of the Chinese Academy of Sciences grant No. XDB23000000, the National Natural Science Foundation of China grant No. 11690024, and by the Astronomical Big Data Joint Research Center, co-founded by the National Astronomical Observatories, Chinese Academy of Sciences and the Alibaba Cloud. 
Portions of this work performed at NRL are supported by the Chief of Naval Research.
The Flatiron Institute is supported by the Simons Foundation.

We thank Logan O'Beirne for helping to uncover a bug in an early implementation of the search algorithms, 
and for providing comments on the algorithms and manuscript. 
We thank Casey McGrath for performing some of the CW analyses. 
We thank Tingting Liu for helpful discussions. 
We thank Alberto Sesana, Xinjiang Zhu, and Siyuan Chen for providing useful comments on the manuscript.

We are grateful for computational resources provided by the Leonard E Parker
Center for Gravitation, Cosmology and Astrophysics at the University of
Wisconsin-Milwaukee, which is supported by NSF Grants 0923409 and 1626190. 
This research made use of the Super Computing System (Spruce Knob) at WVU, 
which is funded in part by the National Science Foundation EPSCoR 
Research Infrastructure Improvement Cooperative Agreement \#1003907, the 
state of West Virginia (WVEPSCoR via the Higher Education Policy Commission)
and WVU.
This research also made use of the Illinois Campus Cluster, a computing resource that is operated by the Illinois Campus Cluster Program (ICCP) in conjunction with the National Center for Supercomputing Applications (NCSA) and which is supported by funds from the University of Illinois at Urbana-Champaign.
Data for this project were collected using the facilities of the Green Bank Observatory and the Arecibo Observatory.
The Green Bank Observatory is a facility of the National Science Foundation 
operated under cooperative agreement by Associated Universities, Inc. 
The Arecibo Observatory is a facility of the National Science Foundation 
operated under cooperative agreement by the University of Central Florida 
in alliance with Yang Enterprises, Inc. and Universidad Metropolitana.

\textit{Software:} \texttt{enterprise} \citep{enterprise}, \texttt{PAL2} \citep{evh17a}, \texttt{PTMCMCSampler} \citep{evh17b}

\appendix

\section{Jump proposals from empirical distributions}
\label{sec:appendix}

In our data analysis pipelines, we computed the posterior distributions 
using MCMC algorithms, which explore parameter space 
through a random walk. 
The CW search for the 11-year data set included 154 parameters: 
7 common GW parameters, 
68 pulsar-term GW parameters, 
68 pulsar red noise parameters, 
and 11 \bayesephem\ parameters. 
Exploring such a large parameter space is computationally intensive, 
and many iterations are required to burn-in the parameters 
and ensure the chains have converged. 
Jump proposals determine how proposed samples are generated, 
and using particularly good jump proposals can significantly decrease the burn-in and convergence time. 
Appendix C of \citetalias{abb+14} discusses the jump proposals used in the 
CW search of the NANOGrav 5-year data set, 
which were also used in the analyses described in this paper.

In the course of analyzing the 11-year data set, 
we introduced a new type of jump proposal 
for the pulsars' red noise parameters and the \bayesephem\ parameters. 
These jump proposals chose new parameter values 
by drawing from empirical distributions 
based on the posteriors from an initial Bayesian analysis 
searching over all of the pulsars that included only 
the pulsars' red noise and \bayesephem; i.e. excluding any GW term. 
These jump proposals do not alter the likelihood or the priors -- 
they ensure that the sampler is choosing new parameters that have a high probability 
of improving the fit, but they do not affect 
the probability that the new parameter values will be accepted or rejected.

This initial pilot run included only 79 parameters, 
and therefore the red noise and \bayesephem\ parameters burned-in relatively quickly.
We constructed the empirical distributions from histograms of the posteriors, 
adding one sample to all bins so that the probability density function was nonzero at every point in the prior. 
For the red noise parameters, we constructed 2-dimensional empirical distributions 
for the amplitude $\log_{10} A$ and spectral index $\gamma$ for each pulsar. 
For the \bayesephem\ parameters, we constructed 1-dimensional empirical distributions 
for each of the six Jupiter orbital elements, which describe perturbations to Jupiter's orbit.

We have found that including jumps that draw from the empirical distributions to the MCMC 
dramatically reduces the number of samples needed for the chains to burn-in and converge 
because the red-noise and \bayesephem\ parameters converge almost immediately. 
Efficiently sampling the pulsars' red noise parameters will become increasingly important 
as the number of pulsars in our PTA increase, 
as each pulsar added to the PTA adds two red noise parameters 
and two pulsar-term parameters to the model.

\bibliographystyle{../style_files/yahapj}
\bibliography{../style_files/apjjabb,bib}

\end{document}